\documentclass{article}
\usepackage{spconf,amsmath,graphicx}

\usepackage{cite,xcolor}
\usepackage{booktabs} 
\usepackage{multirow}
\usepackage{multicol}
\usepackage{amsfonts}

\title{Muse: Multi-modal target speaker extraction with visual cues}
\name{Zexu Pan$^{1,2}$, Ruijie Tao$^{3}$, Chenglin Xu$^{3}$,  Haizhou Li$^{1,3,4}$ \thanks{This research is supported by Programmatic Grant No. A18A2b0046 from the Singapore Government’s Research, Innovation and Enterprise 2020 plan (Advanced Manufacturing and Engineering domain); Human-Robot Interaction Phase 1 (Grant No. 192 25 00054), the National Research Foundation, Prime Minister’s Office, Singapore under the National Robotics Programme. This research is also funded by the Deutsche Forschungsgemeinschaft (DFG, German Research Foundation) under Germany's Excellence Strategy (University Allowance, EXC 2077, University of Bremen).}}

\address{
  $^1$Institute of Data Science, National University of Singapore (NUS), Singapore\\
  $^2$Graduate School for Integrative Sciences and Engineering, NUS, Singapore\\
  $^{3}$Department of Electrical and Computer Engineering, NUS, Singapore\\
  $^{4}$Machine Listening Lab, University of Bremen, Germany}
\begin{document}
\ninept
\maketitle
\begin{abstract}
Speaker extraction algorithm relies on the speech sample from the target speaker as the reference point to focus its attention. Such a reference speech is typically pre-recorded. On the other hand, the temporal synchronization between speech and lip movement also serves as an informative cue. Motivated by this idea, we study a novel technique to use speech-lip visual cues to extract reference target speech directly from mixture speech during inference time, without the need of pre-recorded reference speech. We propose a multi-modal speaker extraction network, named MuSE, that is conditioned only on a lip image sequence. MuSE not only outperforms other competitive baselines in terms of SI-SDR and PESQ, but also shows consistent improvement in cross-dataset evaluations.
\end{abstract}

\section{Introduction}
\label{sec:intro}
Speech separation seeks to separate speakers' voices from a multi-talk acoustic environment, also known as the cocktail party problem~\cite{bronkhorst2000cocktail}. This is a non-trivial task for machines but useful for downstream applications such as speaker verification and speech recognition. Recent advances include computational auditory scene analysis~\cite{hu2007auditory}, non-negative matrix factorization~\cite{virtanen2007monaural}, Deep clustering~\cite{hershey2016deep}, ConvTasnet~\cite{luo2019conv}, and cuPIT~\cite{xu2018single}. However, the formulation of speech separation usually requires the number of speakers to be known in advance, which is not practical in real-world applications.

Humans have the ability to focus the auditory attention to one of the sound sources in a complex cocktail party~\cite{Chenglin2020spex}. Speaker extraction is the technique to emulate such human ability. It generally requires an auxiliary reference that identifies the target speaker. The reference speech can be considered as an attractor for selective auditory attention. It is usually encoded in an utterance-level speaker embedding as speaker's voice signature~\cite{wang2019voicefilter,Chenglin2020spex,he2020speakerfilter}. One example is VoiceFilter that estimates a speaker-conditioned spectrogram mask for target speaker extraction~\cite{wang2019voicefilter}; SpEx/SpEx+ is another successful implementation that trains speaker embedding network jointly with speaker extraction network~\cite{Chenglin2020spex,spex_plus2020}. 

Visual cues also provide informative reference points for target speakers, such as the face or lip image sequences, where phonetic or prosodic content coincides with muscle movement~\cite{ephrat2018looking,afouras2018conversation,wu2019time}. For example, {Ephrat et al.} uses synchronized face embeddings for speaker extraction~\cite{ephrat2018looking}, which leverages the temporal correlation between facial expression and speech. A time-domain audio-visual speech separation network, referred to as AV-ConvTasnet in this paper, is another example, which uses pre-trained lip embedding from lip-reading task for speaker extraction~\cite{wu2019time}. The use of visual cues is intuitively motivated as they are neither corrupted by acoustic noise nor reverberation, furthermore, viseme sequence and phoneme sequence are temporally correlated. However, these methods do not directly use speaker's voice signature as the attractor, which was proven effective in most of the speaker extraction systems.

Visual cues and speaker's voice signature provide complementary information. SpeakerBeam is a technique to leverage on both for speaker extraction~\cite{ochiai2019multimodal}. But it requires pre-registration of reference speech in advance, which limits the scope of real-world applications. For example, a kiosk or service robot in a public area doesn't have the privilege to pre-register visitors or passers-by. This prompts us to look into speaker extraction scenarios~\cite{roth2020ava} where audio-visual stream of data are available, but speech or face pre-registration of the target speaker is not.

A previous study~\cite{Afouras19b} suggested a two-pass mechanism to self-enroll the speaker's voice signature, which aims to address the visual occlusion problem. In the first pass, it extracts the target speech with only the visual cues. In the second pass, it derives the target speaker embedding from the output of the first pass, which is used as the attractor to deal with the situation where visual cues are occluded. The use of self-enrolled speaker embedding and visual cues in the two-pass system outperforms the one-pass system with only visual cues in face of 80\% occlusion. However, there is no improvement for the situation where the visual cues are always available. One possible reason could be that the two-pass are not jointly optimized. Furthermore, the two-pass mechanism is complex and computationally heavy.

\begin{figure*}[htb]
    \begin{minipage}[b]{.4\linewidth}
      \centering
      \centerline{\includegraphics[height=5cm]{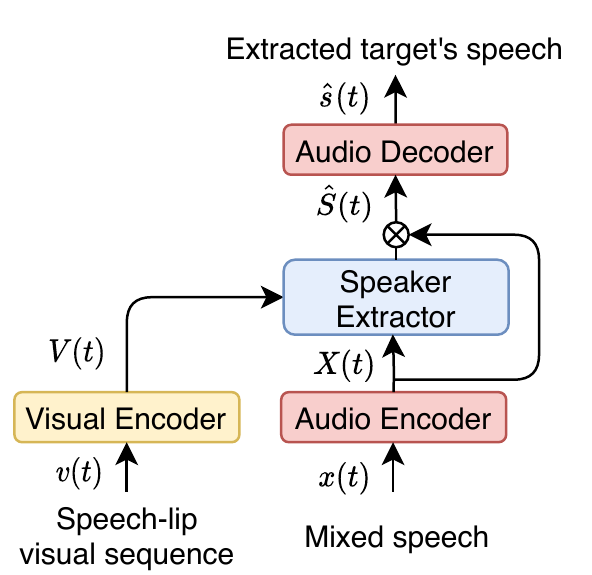}}
      \centerline{(a) Overall structure of MuSE}\medskip
    \end{minipage}
    \hfill
    \begin{minipage}[b]{0.26\linewidth}
      \centering
      \centerline{\includegraphics[height=5cm]{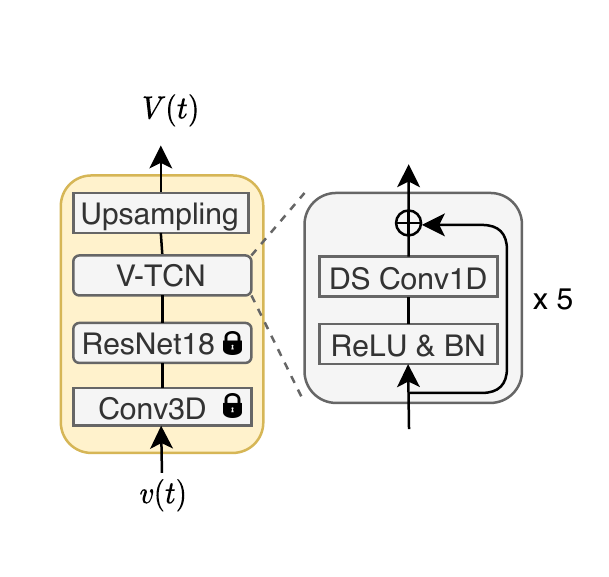}}
      \centerline{(b) Visual encoder}\medskip
    \end{minipage}
    \hfill
    \begin{minipage}[b]{0.26\linewidth}
      \centering
      \centerline{\includegraphics[height=5cm]{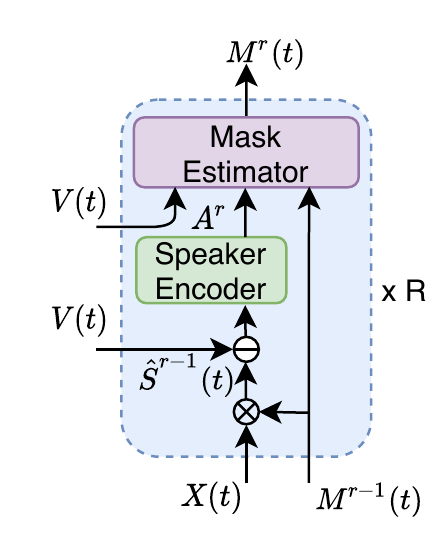}}
      \centerline{(c) Speaker extractor}\medskip
    \end{minipage}
    \vspace{-0.3cm}
    \caption{(a) MuSE is a mask-based structure, consisting of a visual encoder, an audio encoder, a speaker extractor, and a decoder.  MuSE seeks to extract a speaker's voice through masking similar to~\cite{wu2019time}, (b) Visual encoder,  (c) The $r^{th}$ extractor block in speaker extractor. $\oplus$, $\otimes$ refer to point-wise addition and multiplication, $\ominus$ refers to concatenation over temporal dimension.}
    \label{fig:overall}
    \vspace{-5pt}
\end{figure*}

We propose a multi-modal speaker extraction network (MuSE) that uses speech-lip synchronization cue to extract self-enrolled speaker embedding representing the target speaker's voice signature. In this way, neither speaker pre-registration nor reference speech is required. MuSE then uses both speech-lip visual cues and self-enrolled speaker embedding as the condition to predict target speech.
Unlike the two-pass system\cite{Afouras19b}, our work gradually improves the qualities of extracted speech and self-enrolled speaker embedding together with the visual cues in a jointly optimized framework. MuSE not only outperforms the baseline in terms of signal and perceptual quality but also shows consistent improvement on cross-dataset evaluations.

\section{MuSE network}
\label{sec:proposed_method}

\subsection{MuSE vs AV-ConvTasnet}
MuSE adopts the mask-based speaker extraction mechanism and shares a similar overall network architecture with AV-ConvTasnet, which consist of four parts~\cite{wu2019time}, namely audio encoder, audio decoder, visual encoder, and speaker extractor as shown in Fig.~\ref{fig:overall}(a). 

Visual encoder encodes video stream $v(t)$ into a sequence of visual embeddings $V(t)$. Audio encoder transforms the input speech mixture $x(t)$ into latent representation $X(t)$, that is also referred to as speech embedding, while audio decoder renders the extracted speech $\hat{s}(t)$ from its latent representation $\hat{S}(t)$. In AV-ConvTasnet, speaker extractor takes $V(t)$ and $X(t)$ as input, and estimates the mask for target's speech $M^r(t)$ with $V(t)$ as reference; while in MuSE, speaker extractor performs a similar task, but with a more complex architecture, that is illustrated in Fig.~\ref{fig:overall}(c) and discussed in section 2.4. The MuSE speaker extractor represents the main contributions of this work.


\subsection{Audio encoder/decoder}
The audio encoder performs $1$D convolution on $x(t) \in \mathbb{R}^{1 \times T}$ with a kernel size $L$ and stride $L/2$.
\begin{equation}
    \label{eqa:encoder}
    X(t) = Conv1D(x(t), L, L/2)  \;  \in \mathbb{R}^{N\times K}
\end{equation}
where $N$ is the speech embedding dimension and $K = (2(T-L))/L+1$. As $x(t)$ is a time-domain signal, the 1D convolution behaves like a frequency analyzer~\cite{luo2019conv}. The audio decoder performs overlap and add operation to reconstruct $\hat{s}(t)$ from $\hat{S}(t)$~\cite{oppenheim1978theory}. 
\begin{equation}
    \label{eqa: decoder}
    \hat{s}(t) = Overlap \& Add(\hat{S}(t), L, L/2)  \;  \in \mathbb{R}^{1\times T}
\end{equation}
\begin{equation} 
    \hat{S}(t) = X(t) \otimes M_r(t)  \;  \in \mathbb{R}^{N\times K}
\end{equation}
$N$ and $L$ are set to $256$ and $40$ respectively in this paper.

\subsection{Visual encoder}
As shown in Fig.~\ref{fig:overall}(b), the visual encoder has a 3D convolutional layer (Conv3D) followed by a ResNet18 block~\cite{afouras2018deep}, and an additional video temporal convolutional block (V-TCN) consists of $5$ residual connected rectified linear unit (ReLU), batch normalization (BN) and depth-wise separable convolutional layers (DS Conv1D)~\cite{wu2019time}. The inputs to the visual encoder are cropped lip images synchronized with speech. Each image is encoded into a single vector embedding of size $512$. The output of V-TCN is up-sampled to match the temporal resolution of speech embedding. Denoted with a lock in the figure, the Conv3D and ResNet18 have the same network architecture and are pre-trained from the lip-reading task according to work~\cite{afouras2018deep}, their weights are fixed during the speaker extraction training. With the visual encoder, we seek to retain the lip-reading ability and encode the viseme movements that synchronizes phoneme sequence of speech.

\subsection{Speaker extractor}
\label{sec:speaker_extractor}
Iterated mask estimator progressively improves the quality of extracted speech~\cite{luo2019conv, Chenglin2020spex}. SpEx encodes the pre-enrolled utterance and feeds the speaker embedding into each mask estimator~\cite{Chenglin2020spex}. Encouraged by the positive results, we design our speaker extractor in an iterative manner, consisting of a stack of $R$ extractor blocks, each of which consists of a speaker encoder and a mask estimator.

The speaker encoder encodes the on-the-fly estimated speech signals into audio cues $A^r$. As such on-the-fly estimated speech signals are of varying qualities before each mask estimator, we hypothesize that it is more logical for each speaker encoder to have its own weights (as opposed to the shared weights in SpEx) for each mask estimator. The mask estimator estimates a mask $M^r(t)$ that only lets pass the target speech. $A^r$ is a fixed dimension embedding vector, that encodes an audio-visual sequence of variable length, while $M^r(t)$ is a sequence of masking frames, each corresponds to an input speech frame. They are progressively refined through $R$ repeated blocks. The $r^{th}$ block is shown in Fig.~\ref{fig:overall}(c), with $M^0(t) = X(t)$. We choose $R=4$ in this paper.

\begin{figure}[htb]
\begin{minipage}[b]{.49\linewidth}
  \centering
  \centerline{\includegraphics[height=5cm]{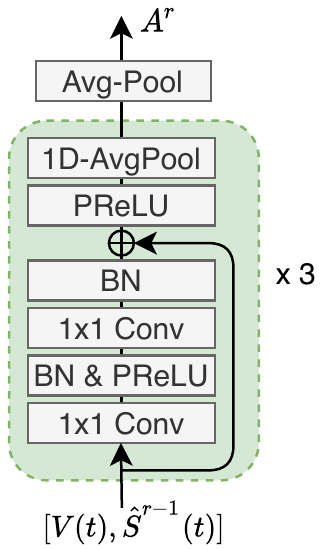}}
  \centerline{(a) Speaker encoder}\medskip
\end{minipage}
\hfill
\begin{minipage}[b]{0.49\linewidth}
  \centering
  \centerline{\includegraphics[height=5cm]{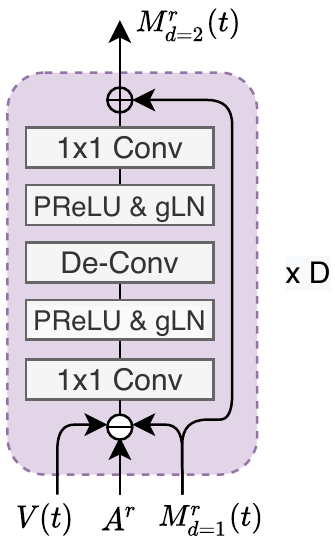}}
  \centerline{(b) Mask estimator}\medskip
\end{minipage}
\vspace{-0.5cm}
\caption{Architecture and data flow of speaker encoder and mask estimator. All point-wise convolutional layers (1x1 Conv) have kernel size of $1$. In speaker encoder, the channel size of 1x1 Conv are set to $256$. In mask estimator, the first and last 1x1 Conv have kernel size of $512$ and $256$ respectively; the depth-wise convolutional layer (De-Conv) have kernel size $3$ and channel size $512$. BN, global layer normalization (gLN) and parametric ReLU (PReLU) are used as activation and normalization functions.}
\label{fig:detail}
\vspace{-10pt}
\end{figure}

\subsubsection{Speaker encoder}
As illustrated in Fig.~\ref{fig:detail}(a), speaker encoder consists a stack of three residual blocks followed by an adaptive average pooling layer (Avg-Pool)~\cite{spex_plus2020}. The speaker encoder takes a temporal sequence $V(t) \ominus \hat{S}^{r-1}(t)$ as input, where
\begin{equation}
    \hat{S}^{r-1}(t) = X(t) \otimes M^{r-1}(t)
\end{equation}
is the masked speech embedding for target speech from the $(r-1)^{th}$ extractor block. The 1D-AvgPool layer has kernel size of $3$, while the Avg-Pool layer pool the temporally spanned speaker embeddings into a single vector $A^r$ of dimension $256$, representing the speaker encoded at extractor block $r$. 

\subsubsection{Mask estimator}
As shown in Fig.~\ref{fig:detail}(b), the mask estimator shares the same network architecture as AV-ConvTasnet~\cite{wu2019time}. It consists of a stack of $D$ audio temporal convolution block (A-TCN) with an exponential growth dilation factor $2^d$ in the De-Conv. We set $D=8$ here. The input to the first A-TCN in the stack includes a sequence of visual embeddings $V(t)$, a fixed dimension target speaker embedding $A^r$, and a sequence of masking frames $M^r_{d=1}(t)$, while to the other A-TCNs are $M^r_d(t)$ only.

\subsection{Multi-task learning}
To ensure MuSE optimizes both the discriminative speaker embedding and the target speaker's voice, we propose a multi-task learning framework with two objectives. A scale-invariant signal-to-noise ratio (SI-SDR)~\cite{le2019sdr} loss is to measure the quality between the extracted and clean target speech. The cross-entropy (CE) loss is used for speaker classification. The CE loss is applied to every speaker extractor network in the $R$ repeats of the extractor block. The overall loss is defined in equation \ref{eqa:loss_all}.
\begin{equation}
    \label{eqa:loss_all}
    \mathcal{L} = \mathcal{L}_{SI\mbox{-}SDR} + \gamma \mathcal{L}_{CE}
\end{equation}
\begin{equation}
    \label{eqa:loss_sisnr}
    \mathcal{L}_{SI\mbox{-}SDR} = - 20 \log_{10} \frac{||\frac{<\hat{s},s>s}{||s||^2}||}{||\hat{s} - \frac{<\hat{s},s>s}{||s||^2}||}
\end{equation}
\begin{equation}
    \label{eqa:loss_ce}
    \mathcal{L}_{CE} = - \sum_{r=1}^{R}\sum_{c=1}^{C}y_c \log (softmax(W^rA^r))
\end{equation}
where $\gamma$ is a scaling factor and set to $0.1$. $s$ is the ground truth target speech,  $y$ is target speaker's class label respectively. $C$ is the number of speakers in the dataset. $W^r$ is a learnable weight matrix in each output layer for speaker classification.

\section{Experiment setup}
\label{sec:experiment}

We develop a MuSE system as described in Section 2 \footnote{The code is available at https://github.com/zexupan/MuSE.}, and re-implement the AV-ConvTasnet as our baseline~\cite{wu2019time} for comparison.
We first evaluate on the VoxCeleb2 dataset~\cite{chung2018voxceleb2}. The speakers in the train set and test set do not overlap. All utterances have more than $4$ seconds of duration.  We obtain $48,000$ utterances from 800 speakers in the train set, and $36,237$ utterances from 118 speakers in the test set, we create $2$ and $3$ speakers mixtures of $20,000$, $5,000$, $3,000$ utterances for train, validation and test sets, respectively. Interference speech is mixed with the target speech with a random Signal-to-Noise (SNR) ratio between $10$dB to $-10$dB. The videos are sampled at $25$ FPS. The audio is synchronized with the video and sampled at $16$ kHz.

We also compare MuSE and AV-ConvTasnet baseline on cross-domain datasets, namely Grid, TCD-TIMIT, LRS2, LRS3 and AVSpeech datasets~\cite{cooke2006audio,harte2015tcd,afouras2018deep,afouras2018lrs3,ephrat2018looking}. Grid and TCD-TIMIT are studio videos while LRS2, LRS3 and AVSpeech are created from BBC, TED videos and YouTube videos, respectively. These datasets are pre-processed with face detection and tracking same as the VoxCeleb2 dataset to minimize the visual mismatch. We generate $3,000$ utterances to form a test set for each of the above datasets, following VoxCeleb2 protocol. 

In training MuSE, adam optimizer is used with an initial learning rate of $1e^{-3}$. The learning rate is halved if the validation loss increases consecutively for $3$ epochs. The training stops when validation loss increases for 5 epochs. During inference, only $\hat{s}(t)$ is extracted while the estimated speaker's class label in Equation \ref{eqa:loss_ce} is redundant.

\section{Evaluation Results}
\label{sec:result}
\subsection{MuSE vs AV-ConvTasnet on VoxCeleb2}
We report the performance in terms of SI-SDR improvement (SI-SDRi) and PESQ~\cite{rix2001perceptual}. SI-SDRi measures the speech signal quality while PESQ measures the overall perceptual quality. PESQ is on a scale from $1$ to $5$, the higher the better. As shown in Table \ref{table:result_in_domain}, MuSE shows a relative improvement of $1$ dB and $2$ dB in terms of  SI-SDRi over AV-ConvTasnet with $2$ speakers and $3$ speakers mixtures. MuSE also has $7\%$ and $10\%$ relative improvement on PESQ.

\subsection{MuSE vs AV-ConvTasnet for cross-dataset test}
We also evaluated MuSE on other audio-visual datasets. As shown in Table \ref{table:result_cross_domain}, we classify the $5$ datasets into ``Wild" and ``Studio" domains. LRS2, LRS3 and AVSpeech belong to ``Wild" videos similar to VoxCeleb2. Grid and TCD-TIMIT are collected in studio.

On LRS2, LRS3 and AVSpeech datasets, MuSE shows the relative improvements of $1$ dB and $2$ dB on SI-SDRi on $2$ speakers and $3$ speakers mixtures, which are consistent with the improvement on VoxCeleb2 dataset. 
On the Grid and TCD-TIMIT datasets, which belong to another domain, MuSE shows a larger relative improvement on SI-SDRi of $1.5$ dB on two speaker mixtures. The difference between MuSE and AV-ConvTasnet is the use of speaker embedding. The results validate the benefit of using speaker embedding in MuSE, which represents a more direct speaker reference point than speech-lip synchronization cue alone. However, on $3$ speakers mixture, MuSE has a smaller relative gain of about $1$ dB. The could be due to the fact that, as the number of speakers in the mixture increases, the speaker embedding estimated becomes noisy and less representative.

\begin{table}
    \centering
    \caption{MuSE shows consistent improvement over AV-ConvTasnet on VoxCeleb2 dataset.}
    \label{table:result_in_domain}
    \begin{tabular}{c c c c} 
       \toprule
        Dataset         & Model               &  SI-SDRi (dB)      & PESQ\\        
       \midrule
       \multirow{2}*{2 Spkr}
               &        AV-ConvTasnet        &  10.641       & 2.065\\
               &        MuSE                 &  11.673       & 2.211\\
       \midrule
       \multirow{2}*{3 Spkr}
               &        AV-ConvTasnet        & 9.778   &  1.470\\
               &        MuSE                 &  11.643       &1.623 \\
       \bottomrule
    \end{tabular}
    \vspace{-4mm}
\end{table}

\begin{table}
    \centering
    \caption{Cross-datasets evaluations (SI-SDRi in dB) of the baseline and our model. Models are trained on VoxCeleb2 and tested on other datasets.}
    \label{table:result_cross_domain}
    \addtolength{\tabcolsep}{-1pt}
    \begin{tabular}{c c c c c c} 
       \toprule
        Domain   &Dataset         & Model               &  2-mix  &  3-mix\\        
       \midrule
       \multirow{6}*{Wild}    &\multirow{2}*{LRS2}        &AV-ConvTasnet        &  10.81  &9.50\\
                                                        &&MuSE                 &  11.97  &11.13\\ \cline{2-5}
                            &\multirow{2}*{LRS3}        &AV-ConvTasnet        &  12.50  & 10.52\\
                                                        &&MuSE                 &  13.44 & 12.34\\ \cline{2-5}
                            &\multirow{2}*{AVSpeech}    &AV-ConvTasnet        &  4.96  & 2.77\\
                                                        &&MuSE                 &  5.88  & 4.81\\
       \midrule
       \multirow{4}*{Studio}   & \multirow{2}*{Grid}       &AV-ConvTasnet        &  6.84  & 5.53\\
                                                        &&MuSE                 &  8.53  & 6.88\\ \cline{2-5}
                            &\multirow{2}*{TCD-TIMIT}   &AV-ConvTasnet        &  11.35   & 9.60\\
                                                        &&MuSE                 &  12.84  & 10.48\\
       \bottomrule
    \end{tabular}
    \addtolength{\tabcolsep}{1pt}
    \vspace{-2mm}
\end{table}

\subsection{Missing visual cues}
We further examine the effect when visual cues are missing on VoxCeleb2 2-speaker mixtures. A random percentage from $10\%$ to $80\%$ of visual segment is occluded for half of the mixtures during training and evaluation. We set the missing visual input to zero vectors, to simulate the case when the face detection and tracking algorithm fails to find the target person. The AV-ConvTasnet has SI-SDRi of $10.421$ dB while MuSE has SI-SDRi of $11.328$ dB. MuSE shows $0.9$ dB improvement over the baseline. The performance of MuSE only sees a drop of $0.343$ dB over the non-missing scenario.

\subsection{Comparison with related work}
We compare MuSE with recent audio-visual speaker extraction studies on VoxCeleb2 datasets in Table \ref{table:result_compare} in terms of signal-to-distortion ratio improvement (SDRi). MuSE is as competitive as the state-of-the-art method despite that we only used a subset of the VoxCeleb2 dataset for training. We understand that the SDRi results are not directly comparable as they are not trained on exactly the same dataset.

MuSE is different from the 'Looking to listen at the cocktail party'~\cite{ephrat2018looking}. As MuSE uses speech-lip synchronization information instead of speech-face synchronization cue, MuSE is expected to generalize well for new speakers.

\subsection{Ablation studies}
\label{sec:ablation}
We study a variant of MuSE that is only trained with the SI-SDR loss without the speaker CE loss, referred to as MuSE-jt. As shown in Table \ref{table:ablation}, MuSE performs similarly with MuSE-jt on VoxCeleb2 and Grid dataset, but outperforms it on TCD-TIMIT dataset. The difference between MuSE and MuSE-jt is the use of speaker loss as shown in Equation (7). The results suggest that the speaker loss improves the domain robustness of the system.

To prove our hypothesis on the importance of non-shared weights speaker encoder in section~\ref{sec:speaker_extractor}. We study a variant of MuSE with shared-weights speaker encoders. The SI-SDRi drops from $11.673$dB to $11.231$dB. Proving the non-shared speaker encoder is necessary for every iteration.

In addition, we study the extraction performance with different number of iterations $R$. As $R$ increases from $1$ to $4$, the SI-SDRi of the extracted speech increase to $7.248$dB, $10.48$dB, $11.324$dB, $11.673$dB respectively. The speech quality converges as the number of iteration increases.

\begin{table}
    \centering
    \caption{A comparison across various audio-visual speaker extraction implementations on VoxCeleb2 2-speaker mixture, using video (V), a static image (I) or speaker embedding (A) as auxiliary inputs.}
    \label{table:result_compare}
    \begin{tabular}{c c c c} 
       \toprule
        Model                                  &Auxiliary input & SDRi (dB)  \\        
       \midrule
        AVS~\cite{8958547}                     & V (face) +A &5.9  \\
        AV-BLSTM~\cite{ephrat2018looking,8958547}  & V (face)  &  3.25 \\
        FaceFilter~\cite{chung2020facefilter}  & I (face) &2.5  \\
        AV-U-Net~\cite{owens2018audio}         & V (face) & 7.6    \\
        AV-LSTM~\cite{afouras2018conversation}  & V (lips)       & 12.1   \\
       \midrule
        AV-ConvTasnet    & V (lips)    &10.9    \\
        MuSE            & V (lips)     &12.0   \\
       \bottomrule
    \end{tabular}
    \vspace{-4mm}
\end{table}

\begin{table}
    \centering
    \caption{SI-SDRi (dB) of ablation studies on 2-speaker mixture on VoxCeleb2, Grid and TCD-TIMIT datasets. SE refers to speaker embedding. }
    \label{table:ablation}
    \addtolength{\tabcolsep}{-3pt}
    \begin{tabular}{c c c c c c} 
       \toprule
           Model      & w/o & VoxCeleb2  &Grid  & TCD-TIMIT  \\   
       \midrule
        AV-ConvTasnet  & SE   &10.64    & 6.84 &11.35 \\
        MuSE-jt & MTL         &11.61    &8.50  &10.07 \\
        MuSE       &  -      &11.67    &8.53   &12.84\\
       \bottomrule
    \end{tabular}
    \addtolength{\tabcolsep}{3pt}
    \vspace{-2mm}
\end{table}

\section{Conclusion}
\label{sec:conclusion}

In this paper, we study a new way to extract the speaker embedding by using speech-lip synchronization cue. The visual information and extracted target speaker embedding are used as references to extract the target speaker's voice. We show that MuSE has a clear advantage over the baseline in both in-domain and out-of-domain tests. We are considering using additional modalities, such as text, as additional input to the mask estimator.


\vfill\pagebreak

\small
\bibliographystyle{IEEEbib}
\bibliography{mybib}

\begin{thebibliography}{10}

\bibitem{bronkhorst2000cocktail}
Adelbert~W Bronkhorst,
\newblock ``The cocktail party phenomenon: A review of research on speech
  intelligibility in multiple-talker conditions,''
\newblock {\em Acta Acustica united with Acustica}, vol. 86, no. 1, pp.
  117--128, 2000.

\bibitem{hu2007auditory}
Guoning Hu and DeLiang Wang,
\newblock ``Auditory segmentation based on onset and offset analysis,''
\newblock {\em IEEE transactions on audio, speech, and language processing},
  vol. 15, no. 2, pp. 396--405, 2007.

\bibitem{virtanen2007monaural}
Tuomas Virtanen,
\newblock ``Monaural sound source separation by nonnegative matrix
  factorization with temporal continuity and sparseness criteria,''
\newblock {\em IEEE transactions on audio, speech, and language processing},
  vol. 15, no. 3, pp. 1066--1074, 2007.

\bibitem{hershey2016deep}
John~R Hershey, Zhuo Chen, Jonathan Le~Roux, and Shinji Watanabe,
\newblock ``Deep clustering: Discriminative embeddings for segmentation and
  separation,''
\newblock in {\em IEEE International Conference on Acoustics, Speech and Signal
  Processing}. IEEE, 2016, pp. 31--35.

\bibitem{luo2019conv}
Yi~Luo and Nima Mesgarani,
\newblock ``Conv-{TasN}et: Surpassing ideal time--frequency magnitude masking
  for speech separation,''
\newblock {\em IEEE/ACM transactions on audio, speech, and language
  processing}, vol. 27, no. 8, pp. 1256--1266, 2019.

\bibitem{xu2018single}
Chenglin Xu, Wei Rao, Xiong Xiao, Eng~Siong Chng, and Haizhou Li,
\newblock ``Single channel speech separation with constrained utterance level
  permutation invariant training using grid lstm,''
\newblock in {\em IEEE International Conference on Acoustics, Speech and Signal
  Processing}. IEEE, 2018, pp. 6--10.

\bibitem{Chenglin2020spex}
Chenglin Xu, Wei Rao, Eng~Siong Chng, and Haizhou Li,
\newblock ``Sp{E}x: Multi-scale time domain speaker extraction network,''
\newblock {\em IEEE/ACM transactions on audio, speech, and language
  processing}, 2020.

\bibitem{wang2019voicefilter}
Quan Wang, Hannah Muckenhirn, Kevin Wilson, Prashant Sridhar, Zelin Wu, John~R
  Hershey, Rif~A Saurous, Ron~J Weiss, Ye~Jia, and Ignacio~Lopez Moreno,
\newblock ``Voice{F}ilter: Targeted voice separation by speaker-conditioned
  spectrogram masking,''
\newblock {\em Proc. Interspeech}, pp. 2728--2732, 2019.

\bibitem{he2020speakerfilter}
Shulin He, Hao Li, and Xueliang Zhang,
\newblock ``Speakerfilter: Deep learning-based target speaker extraction using
  anchor speech,''
\newblock in {\em IEEE International Conference on Acoustics, Speech and Signal
  Processing}. IEEE, 2020, pp. 376--380.

\bibitem{spex_plus2020}
Meng Ge, Chenglin Xu, Longbiao Wang, Eng~Siong Chng, Jianwu Dang, and Haizhou
  Li,
\newblock ``Sp{E}x+: A complete time domain speaker extraction network,''
\newblock in {\em Proc. Interspeech}, 2020.

\bibitem{ephrat2018looking}
Ariel Ephrat, Inbar Mosseri, Oran Lang, Tali Dekel, Kevin Wilson, Avinatan
  Hassidim, William~T Freeman, and Michael Rubinstein,
\newblock ``Looking to listen at the cocktail party: a speaker-independent
  audio-visual model for speech separation,''
\newblock {\em ACM Transactions on Graphics}, vol. 37, no. 4, pp. 1--11, 2018.

\bibitem{afouras2018conversation}
Triantafyllos Afouras, Joon~Son Chung, and Andrew Zisserman,
\newblock ``The {C}onversation: Deep audio-visual speech enhancement,''
\newblock {\em Proc. Interspeech}, pp. 3244--3248, 2018.

\bibitem{wu2019time}
Jian Wu, Yong Xu, Shi-Xiong Zhang, Lian-Wu Chen, Meng Yu, Lei Xie, and Dong Yu,
\newblock ``Time domain audio visual speech separation,''
\newblock in {\em IEEE Automatic Speech Recognition and Understanding
  Workshop}. IEEE, 2019, pp. 667--673.

\bibitem{ochiai2019multimodal}
Tsubasa Ochiai, Marc Delcroix, Keisuke Kinoshita, Atsunori Ogawa, and Tomohiro
  Nakatani,
\newblock ``Multimodal speakerbeam: Single channel target speech extraction
  with audio-visual speaker clues,''
\newblock {\em Proc. Interspeech}, pp. 2718--2722, 2019.

\bibitem{roth2020ava}
Joseph Roth, Sourish Chaudhuri, Ondrej Klejch, Radhika Marvin, Andrew
  Gallagher, Liat Kaver, Sharadh Ramaswamy, Arkadiusz Stopczynski, Cordelia
  Schmid, Zhonghua Xi, et~al.,
\newblock ``{AVA} active speaker: An audio-visual dataset for active speaker
  detection,''
\newblock in {\em IEEE International Conference on Acoustics, Speech and Signal
  Processing}. IEEE, 2020, pp. 4492--4496.

\bibitem{Afouras19b}
T.~Afouras, J.~S. Chung, and A.~Zisserman,
\newblock ``My lips are concealed: Audio-visual speech enhancement through
  obstructions,''
\newblock in {\em INTERSPEECH}, 2019.

\bibitem{oppenheim1978theory}
AV~Oppenheim and R~Schafer,
\newblock ``Theory and application of digital signal processing,''
\newblock {\em Englewood Cliffs}, 1978.

\bibitem{afouras2018deep}
Triantafyllos Afouras, Joon~Son Chung, Andrew Senior, Oriol Vinyals, and Andrew
  Zisserman,
\newblock ``Deep audio-visual speech recognition,''
\newblock {\em IEEE transactions on pattern analysis and machine intelligence},
  2018.

\bibitem{le2019sdr}
Jonathan Le~Roux, Scott Wisdom, Hakan Erdogan, and John~R Hershey,
\newblock ``{SDR}--half-baked or well done?,''
\newblock in {\em IEEE International Conference on Acoustics, Speech and Signal
  Processing}. IEEE, 2019, pp. 626--630.

\bibitem{chung2018voxceleb2}
Joon~Son Chung, Arsha Nagrani, and Andrew Zisserman,
\newblock ``Vox{C}eleb2: Deep speaker recognition,''
\newblock {\em Proc. Interspeech}, pp. 1086--1090, 2018.

\bibitem{cooke2006audio}
Martin Cooke, Jon Barker, Stuart Cunningham, and Xu~Shao,
\newblock ``An audio-visual corpus for speech perception and automatic speech
  recognition,''
\newblock {\em The Journal of the Acoustical Society of America}, vol. 120, no.
  5, pp. 2421--2424, 2006.

\bibitem{harte2015tcd}
Naomi Harte and Eoin Gillen,
\newblock ``{TCD-TIMIT}: An audio-visual corpus of continuous speech,''
\newblock {\em IEEE Transactions on Multimedia}, vol. 17, no. 5, pp. 603--615,
  2015.

\bibitem{afouras2018lrs3}
Triantafyllos Afouras, Joon~Son Chung, and Andrew Zisserman,
\newblock ``{LRS3-TED}: a large-scale dataset for visual speech recognition,''
\newblock {\em arXiv preprint arXiv:1809.00496}, 2018.

\bibitem{rix2001perceptual}
Antony~W Rix, John~G Beerends, Michael~P Hollier, and Andries~P Hekstra,
\newblock ``Perceptual evaluation of speech quality ({PESQ})-a new method for
  speech quality assessment of telephone networks and codecs,''
\newblock in {\em IEEE International Conference on Acoustics, Speech, and
  Signal Processing. Proceedings}. IEEE, 2001, vol.~2, pp. 749--752.

\bibitem{8958547}
Luo Yiyu, Wang Jing, Wang Xinyao, Wen Liang, and Wang Lizhong,
\newblock ``Audio-visual speech separation using i-vectors,''
\newblock in {\em IEEE 2nd International Conference on Information
  Communication and Signal Processing}, 2019, pp. 276--280.

\bibitem{chung2020facefilter}
Soo-Whan Chung, Soyeon Choe, Joon~Son Chung, and Hong-Goo Kang,
\newblock ``{FaceFilter: Audio-Visual Speech Separation Using Still Images},''
\newblock in {\em Proc. Interspeech 2020}, 2020, pp. 3481--3485.

\bibitem{owens2018audio}
Andrew Owens and Alexei~A Efros,
\newblock ``Audio-visual scene analysis with self-supervised multisensory
  features,''
\newblock in {\em Proceedings of the European Conference on Computer Vision},
  2018, pp. 631--648.

\end{thebibliography}

\end{document}